\documentclass{article} 
\usepackage[dvipsnames]{xcolor}         
\usepackage{iclr2021_conference,times}

\usepackage{amsmath,amsfonts,bm}

\def\eqref#1{equation~\ref{#1}}

\def\1{\bm{1}}

\DeclareMathAlphabet{\mathsfit}{\encodingdefault}{\sfdefault}{m}{sl}
\SetMathAlphabet{\mathsfit}{bold}{\encodingdefault}{\sfdefault}{bx}{n}

\usepackage[utf8]{inputenc} 
\usepackage[T1]{fontenc}    
\usepackage[colorlinks=false]{hyperref}       
\usepackage{url}            
\usepackage{booktabs}       
\usepackage{amsfonts}       
\usepackage{nicefrac}       
\usepackage{microtype}      
\usepackage{graphicx}
\usepackage{subcaption}
\usepackage{booktabs} 
\usepackage{amsmath}

\usepackage{algorithm}
\usepackage[noend]{algpseudocode}
\usepackage{algorithmicx}
\algblockdefx{RepeatUntilTimeout}{EndRepeat}{\textbf{repeat until timeout}}{}
\algblockdefx{RepeatUntil}{EndRepeat}{\textbf{repeat until }}{}
\algtext*{EndRepeat}
\usepackage{float}
\newfloat{algorithm}{t}{}

\algnewcommand\algorithmicinput{\textbf{Input:}}
\algnewcommand\Input{\item[\algorithmicinput]}
\algnewcommand\algorithmicoutput{\textbf{Output:}}
\algnewcommand\Output{\item[\algorithmicoutput]}
\algnewcommand\algorithmicdata{\textbf{Auxiliary Data:}}
\algnewcommand\Data{\item[\algorithmicdata]}
\makeatletter
\algnewcommand{\LineComment}[1]{\Statex \hskip\ALG@thistlm \(\triangleright\) #1}
\makeatother

\usepackage{etoolbox}
\usepackage{tikz}
\usetikzlibrary{tikzmark}
\usetikzlibrary{calc}

\newcommand{\ALGtikzmarkcolor}{lightgray}
\newcommand{\ALGtikzmarkextraindent}{4pt}
\newcommand{\ALGtikzmarkverticaloffsetstart}{-.7ex}
\newcommand{\ALGtikzmarkverticaloffsetend}{-.5ex}
\makeatletter
\newcounter{ALG@tikzmark@tempcnta}

\newcommand\ALG@tikzmark@start{%
    \global\let\ALG@tikzmark@last\ALG@tikzmark@starttext%
    \expandafter\edef\csname ALG@tikzmark@\theALG@nested\endcsname{\theALG@tikzmark@tempcnta}%
    \tikzmark{ALG@tikzmark@start@\csname ALG@tikzmark@\theALG@nested\endcsname}%
    \addtocounter{ALG@tikzmark@tempcnta}{1}%
}

\def\ALG@tikzmark@starttext{start}
\newcommand\ALG@tikzmark@end{%
    \ifx\ALG@tikzmark@last\ALG@tikzmark@starttext
    \else
        \tikzmark{ALG@tikzmark@end@\csname ALG@tikzmark@\theALG@nested\endcsname}%
        \tikz[overlay,remember picture] \draw[\ALGtikzmarkcolor] let \p{S}=($(pic cs:ALG@tikzmark@start@\csname ALG@tikzmark@\theALG@nested\endcsname)+(\ALGtikzmarkextraindent,\ALGtikzmarkverticaloffsetstart)$), \p{E}=($(pic cs:ALG@tikzmark@end@\csname ALG@tikzmark@\theALG@nested\endcsname)+(\ALGtikzmarkextraindent,\ALGtikzmarkverticaloffsetend)$) in (\x{S},\y{S})--(\x{S},\y{E});%
    \fi
    \gdef\ALG@tikzmark@last{end}%
}

\apptocmd{\ALG@beginblock}{\ALG@tikzmark@start}{}{\errmessage{failed to patch}}
\pretocmd{\ALG@endblock}{\ALG@tikzmark@end}{}{\errmessage{failed to patch}}
\makeatother

\algnewcommand{\LeftComment}[1]{\Statex \(\triangleright\) #1}

\algrenewcommand\algorithmicindent{1em}

\algrenewcommand{\alglinenumber}[1]{\color{Black}\footnotesize#1:}
\colorlet{BustleLineColor}{Blue!90}

\newcommand{\N}[1]{\textsc{#1}}

\usepackage{enumitem}
\setitemize{itemsep=2pt,topsep=0pt,parsep=0pt,partopsep=0pt,leftmargin=*}
\setenumerate{itemsep=2pt,topsep=0pt,parsep=0pt,partopsep=0pt,leftmargin=*}

\newcommand{\OR}{\; | \;}
\newcommand{\T}[1]{\texttt{#1}}

\newcommand{\calI}{\mathcal{I}}
\newcommand{\calO}{\mathcal{O}}
\newcommand{\calV}{\mathcal{V}}

\usepackage{listings}
\definecolor{codegreen}{rgb}{0,0.6,0}
\definecolor{codegray}{rgb}{0.5,0.5,0.5}
\definecolor{codepurple}{rgb}{0.58,0,0.82}
\definecolor{backcolour}{rgb}{0.95,0.95,0.92}

\lstdefinestyle{mystyle}{
    backgroundcolor=\color{backcolour},
    commentstyle=\color{codegreen},
    keywordstyle=\color{magenta},
    numberstyle=\tiny\color{codegray},
    stringstyle=\color{codepurple},
    basicstyle=\scriptsize,
    breakatwhitespace=false,
    breaklines=true,
    captionpos=b,
    keepspaces=true,
    numbers=left,
    numbersep=5pt,
    showspaces=false,
    showstringspaces=false,
    showtabs=false,
    tabsize=2,
    aboveskip=0em,
    belowcaptionskip=1em,
    belowskip=0em,
}

\newcommand{\alltrue}{\ensuremath{\mathsf{AllTrue}}}
\newcommand{\allfalse}{\ensuremath{\mathsf{AllFalse}}}
\newcommand{\mixed}{\ensuremath{\mathsf{Mixed}}}
\newcommand{\True}{\ensuremath{\mathsf{True}}}
\newcommand{\False}{\ensuremath{\mathsf{False}}}

\title{BUSTLE: Bottom-Up Program Synthesis \\ Through Learning-Guided Exploration}

\author{%
  Augustus Odena\thanks{Equal Contribution}, Kensen Shi\footnotemark[1], David Bieber, Rishabh Singh, Charles Sutton \& Hanjun Dai \\
  Google Research\\
  \texttt{\{augustusodena,kshi,dbieber,rising,charlessutton,hadai\}@google.com}
}

\iclrfinalcopy 
\begin{document}

\maketitle

\begin{abstract}
Program synthesis is challenging largely because of the difficulty of search in a large space of programs. 
Human programmers routinely tackle the task of writing complex programs by writing sub-programs and then analyzing
their intermediate results to compose them in appropriate ways.
Motivated by this intuition, we present a new synthesis approach that leverages learning to guide a bottom-up search over programs.
In particular, we train a model to prioritize compositions of intermediate values during search conditioned on a given set of input-output examples.
This is a powerful combination because of several emergent properties.
First, in bottom-up search, intermediate programs can be executed, providing semantic information to the neural network.
Second, given the concrete values from those executions, we can exploit rich features based on recent work on property signatures.
Finally, bottom-up search allows the system substantial
flexibility in what order to generate the solution, allowing the synthesizer
to build up a program from multiple smaller sub-programs.
Overall, our empirical evaluation finds that the combination of learning 
and bottom-up search is remarkably effective, even with simple supervised learning approaches.
We demonstrate the effectiveness of our technique on two datasets,
one from the SyGuS competition and one of our own creation.
\end{abstract}

\section{Introduction}

Program synthesis is a longstanding goal of artificial intelligence research
~\citep{mannaw71,summers1977methodology},
but it remains difficult in part because of the challenges of search~\citep{sygus,RISHABHSURVEY}.
The objective in program synthesis is to automatically write a program given
a specification of its intended behavior, and current state of the art methods typically perform some form of search over a space of possible programs. Many different search methods
have been explored in the literature, both with and without learning.
These include search within a version-space algebra \citep{FLASHFILL},
bottom-up enumerative search \citep{TRANSIT}, stochastic search~\citep{stoke}, genetic programming~\citep{koza1994genetic}, reducing the synthesis problem to logical satisfiability
\citep{SKETCH}, beam search with a sequence-to-sequence 
neural network \citep{ROBUSTFILL}, learning to perform premise
selection to guide search \citep{DEEPCODER}, learning to prioritize grammar
rules within top-down search \citep{EUPHONY}, and learned search based
on partial executions \citep{REPL,GARBAGECOLLECTOR,ChenLS19}.

While these approaches have yielded significant progress, 
none of them completely capture the 
following important intuition: human programmers routinely 
write complex programs by first writing sub-programs and then analyzing
their intermediate results to compose them in appropriate ways.
We propose a new learning-guided system for synthesis,
called \N{Bustle},\footnote{\underline{\textbf{B}}ottom-\underline{\textbf{U}}p program \underline{\textbf{S}}ynthesis \underline{\textbf{T}}hrough \underline{\textbf{L}}earning-guided \underline{\textbf{E}}xploration} which follows this intuition in
a straightforward manner.
Given a specification of a program's intended behavior
(in this paper given by input-output examples),
\N{Bustle} performs bottom-up enumerative search for a 
satisfying program, following \cite{TRANSIT}.
Each program explored during the bottom-up search
is an expression that
can be executed on the inputs, so we apply a machine learning
model to the resulting value to guide the search.
The model is simply a classifier
trained to predict whether the intermediate value
produced by a partial
program is part of an eventual solution.
This combination of learning and bottom-up search
has several key advantages.
First, because the input to the model is a value produced
by executing a partial program, the model's predictions
can depend on semantic information about the program.
Second, because the search is bottom-up, compared
to previous work on execution-guided synthesis, the search
procedure has more flexibility about which order to generate the
program in, and this flexibility can be exploited by machine learning.

A fundamental challenge in this approach is that
exponentially many intermediate programs are explored during search, 
so the model needs to be both fast and accurate
to yield wall-clock time speedups.
We are allowed to incur some slowdown from performing model inference, because if the model is accurate enough,
we can search many fewer values before finding solutions. However, in the domains
we consider,
executing
a program is still orders of magnitude faster than performing inference on 
even a small machine learning model, so this challenge cannot be ignored.
We employ two techniques to deal with this.
First, we arrange both the synthesizer and the model so that we can batch
model prediction across hundreds of intermediate values. 
Second, we process intermediate expressions using property signatures
\citep{SIGNATURES}, which featurize program inputs and outputs
using another set of programs.

A second challenge is that neural networks require large amounts of data to 
train, but there is no available data source of intermediate expressions.
We can generate programs at random to train the model following
previous work \citep{DEEPCODER,ROBUSTFILL}, but
models trained on random programs do not always 
transfer to human-written benchmarks \citep{shin2019-sv}.
We show that our use of property signatures helps 
with this distribution mismatch problem as well.

In summary, this paper makes the following contributions: 

\begin{itemize}
\item We present \N{Bustle}, which integrates machine learning into bottom-up program synthesis.
\item We show how to efficiently add machine learning 
in the synthesis loop using property signatures and batched predictions. With these techniques, adding the model to the synthesizer provides an end-to-end improvement in synthesis time.
\item 
We evaluate \N{Bustle} on two string transformation datasets: one of our 
own design and one from the SyGuS competition.
We show that \N{Bustle} leads to improvements 
in synthesis time compared to a baseline synthesizer
without learning, a DeepCoder-style synthesizer \citep{DEEPCODER},
and an encoder-decoder model \citep{ROBUSTFILL}.
Even though our model is trained on random programs,
we show that its performance transfers
to a set of human-written synthesis benchmarks.
\end{itemize}

\section{Background and Setup}

\subsection{Programming by Example}

In a Programming-by-Example (PBE) task \citep{PBE,MLPBE,FLASHFILL},
we are given a set of input-output pairs and the goal is to find a program such that for each pair,
the synthesized program generates the corresponding output when executed on the input.
To restrict the search space, the programs are typically restricted to a small domain-specific
language (DSL).
As an example PBE specification, consider the ``io\_pairs''
given in Listing~\ref{listing:signatures}.

\subsection{Our String Transformation DSL}

Following previous work \citep{FLASHFILL,ROBUSTFILL}, we deal with string and
number transformations commonly used in spreadsheets.
Such transformations sit at a nice point on the complexity spectrum 
as a benchmark task;
they are simpler than programs in general purpose languages,
but still expressive enough for many common string transformation tasks. 

The domain-specific language we use (shown in Figure~\ref{fig:sheets_dsl}) is broadly similar to those of
\cite{parisotto2017neuro} and \cite{ROBUSTFILL},
but compared to these, our DSL is expanded in several ways that make the synthesis task more difficult.
First, in addition to string manipulation, our DSL includes integers, 
integer arithmetic, booleans, and conditionals.
Second, our DSL allows for arbitrarily nested expressions,
rather than having a maximum size. 
Finally, and most importantly, previous works~\citep{FLASHFILL, ROBUSTFILL} impose a restriction of having \T{Concat} as the top-level operation. With this constraint, such approaches use version space algebras or dynamic programming to exploit the property that partial programs must form substrings of the output. Our DSL lifts this constraint, allowing the synthesizer to handle more kinds of tasks than in those previous works.

Our DSL allows for compositions of common string transformation functions. These functions include string concatenation (\T{Concat})
and other familiar string operations (listed in Figure~\ref{fig:sheets_dsl} and discussed further in Appendix~\ref{app:dsl}).
Integer functions include arithmetic, finding the index of substrings (\T{Find}), and string length.
Finally, commonly useful string and integer constants
are included. We also use heuristics to extract string constants that appear multiple times in the input-output examples.

\begin{figure}[t]
\small
\begin{alignat*}{2}
\mbox{Expression } E &:= \: && S \OR I \OR B \\
\mbox{String expression } S &:= && \T{Concat}(S_1, S_2) \OR \T{Left}(S, I) \OR \T{Right}(S, I) \OR \T{Substr}(S, I_1, I_2) \\
& && 
\OR \T{Replace}(S_1, I_1, I_2, S_2) \OR \T{Trim}(S) \OR \T{Repeat}(S, I) \OR \T{Substitute}(S_1, S_2, S_3)
\\ & && \OR \T{Substitute}(S_1, S_2, S_3, I) \OR \T{ToText}(I) \OR  \T{LowerCase}(S) \OR \T{UpperCase}(S)
\\ & && \OR \T{ProperCase}(S) \OR T \OR X \OR \T{If}(B, S_1, S_2)\\
\mbox{Integer expression } I &:= && I_1 + I_2 \OR I_1 - I_2 \OR \T{Find}(S_1, S_2) \OR \T{Find}(S_1, S_2, I) \OR \T{Len}(S) \OR J  \\
\mbox{Boolean expression } B &:= && \T{Equals}(S_1, S_2) \OR  \T{GreaterThan}(I_1, I_2) \OR \T{GreaterThanOrEqualTo}(I_1, I_2)\\
\mbox{String constants } T &:= && \texttt{""} \OR \texttt{" "} \OR \texttt{","} \OR \texttt{"."} \OR \texttt{"!"} \OR \texttt{"?"} \OR \texttt{"("} \OR \texttt{")"} \OR \texttt{"["} \OR \texttt{"]"} \OR \texttt{"<"} \OR \texttt{">"} \\
& && \OR \texttt{"\{"} \OR \texttt{"\}"} \OR \texttt{"-"} \OR \texttt{"+"} \OR \texttt{"\_"} \OR \texttt{"/"} \OR \texttt{"\$"} \OR \texttt{"\#"} \OR \texttt{":"} \OR \texttt{";"} \OR \texttt{"@"} \OR \texttt{"\%"} \OR \texttt{"0"} \\
& && \OR \text{string constants extracted from I/O examples} \\
\mbox{Integer constants } J &:= &&  0 \OR 1 \OR 2 \OR 3 \OR 99 \\
\mbox{Input } X &:= && x_1 \OR \ldots \OR x_k
\end{alignat*}
    \caption{Domain-specific language (DSL) of expressions considered in this paper.} 
    \label{fig:sheets_dsl}
\end{figure}

\subsection{Bottom-Up Synthesis}

The baseline synthesizer on top of which we build our approach is a
bottom-up enumerative search inspired by \cite{TRANSIT},
which enumerates DSL expressions from smallest to largest, following Algorithm~\ref{alg:search} if the lines 
colored in blue (\ref{alg:bustle_line_1}, \ref{alg:bustle_line_2}, 
and \ref{alg:bustle_line_3}) are removed.
This baseline uses a \emph{value-based} search.
During the search each candidate expression is executed to see if it meets
the specification. 
Then, rather than storing the expressions that have been produced
in the search, we store the values produced by executing the expressions. 
This allows the search to avoid separately extending sub-expressions that 
are semantically equivalent on the given examples. If there are $n$ separate input-output examples for a given task, each value represents one code expression and internally contains the results of executing that expression on inputs from all $n$ examples. Hence, two values that represent different code expressions are semantically equivalent on the examples if their $n$ contained results all match each other.

Every expression has an integer weight, which for the baseline
synthesizer is the number of nodes in the abstract syntax tree (AST).
The search maintains a table mapping weights to a list of all the values of previously explored sub-expressions of that weight.
The search is initialized with the set of input variables as well as any constants extracted with heuristics, all of which have weight $1$. Extracted constants include common symbols and delimiters that appear at least once, and long substrings that appear multiple times in the example strings. The search then proceeds to create all expressions of weight 2, and then of weight 3, and so on.
To create all values of a particular weight, we loop over all available functions, calling each function with all combinations of arguments that would yield the 
desired weight.
For example, if we are trying to construct all values of weight 10 of the form 
\T{Concat}$(x, y)$, we iterate over all values where $x$ has weight $1$ and $y$ has weight $8$, 
and then  where $x$ has weight $2$ and $y$ has weight $7$, and so forth. 
(The \T{Concat} function itself contributes weight $1$.)
Every time a new expression is constructed, we evaluate it on the given inputs, 
terminating the search when the expression
produces the desired outputs.

\subsection{Property Signatures}
\label{section:property_signatures}

In order to perform machine learning on values encountered during the 
enumeration, we make use of recent work on property signatures 
\citep{SIGNATURES}.
Consider a function with input type $\tau_{in}$ and output type
$\tau_{out}$.
In this context, a property is a function of type:
$(\tau_{in}, \tau_{out}) \rightarrow \texttt{Bool}$ that describes
some aspect of the function under consideration.
If we have a list of such properties and some inputs and outputs of the correct
type, we can evaluate all the properties on the input-output pairs
to get a list of outputs that we will call the property signature.
More precisely, given a list of $n$ input-output pairs and a list of $k$
properties, the property signature is a length $k$ vector.
The elements of the vector corresponding to a given property will have one of
the values \alltrue, \allfalse, and \mixed, depending on whether the property
returned $\True$ for all $n$ pairs, $\False$ for all $n$ pairs, or $\True$ for
some and $\False$ for others, respectively.
Concretely, then, any property signature can be identified with a trit-vector,
and we represent them in computer programs as arrays containing the
values $\{-1, 0, 1\}$. An example of property signatures is shown
in Listing~\ref{listing:signatures}.

\begin{lstlisting}[float=*, language=Python, caption={
An example set of input-output pairs, along with
three properties that can act on them.
The first returns \True\ for the first two pairs and \False\ for the third.
The second returns \False\ for all pairs.
The third returns \True\ for all pairs.
The resulting property signature
is $\{\mixed, \allfalse, \alltrue\}$.
These examples are written in Python for clarity, but our
implementation is in Java.
}, label=listing:signatures]
io_pairs = [
  ("butter",  "butterfly"),
  ("abc",     "abc_"),
  ("xyz",     "XYZ_"),
]

p1 = lambda inp, outp: inp in outp 
p2 = lambda inp, outp: outp.endswith(inp)
p3 = lambda inp, outp: inp.lower() in outp.lower()
\end{lstlisting}

\begin{algorithm}
    \caption{The \N{Bustle} Synthesis Algorithm}
    \label{alg:search}
    \begin{algorithmic}[1]
        \Input Input-output examples $(\mathcal{I}, \mathcal{O})$
        \Output A program $P$ consistent with the examples $(\mathcal{I}, \mathcal{O})$
        \Data Supported operations \emph{Ops}, supported properties \emph{Props}, and a model $M$ trained using \emph{Props} as described in Section \ref{section:training_the_model}.
        \State $E \gets \emptyset $ \Comment{$E$ maps integer weights to terms with that weight}
        \State $C \gets $\Call{ExtractConstants}{$\mathcal{I}, \mathcal{O}$}
        \State $E[1] \gets \mathcal{I} \cup C$ \Comment{Inputs and constants have weight 1}
        \color{BustleLineColor}
        \State $s_{io} \gets \Call{PropertySignature}{\mathcal{I}, \mathcal{O}, \mathit{Props}}$ \label{alg:bustle_line_1}
        \color{Black}
        \For{$w = 2, \dots$} \Comment{Loop over all possible term weights}
            \ForAll{$op \in \mathit{Ops}$}
                \State $n \gets op.\mathit{arity}$
                \State $A \gets \emptyset$  \Comment{$A$ holds all argument tuples}
                \ForAll{$[w_1, \dots, w_n]$ \Comment{Make all arg tuples with these weights that type-check} \\ \hskip\algorithmicindent\hskip\algorithmicindent \hphantom{\textbf{for all}} s.t. $\sum_i w_i = w-1, \enskip w_i \in \mathbb{Z}^+$}
                        \State $A \gets A \cup \{(a_1, \dots , a_n) \mid a_i.\mathit{weight} = w_i \wedge a_i.type = op.argtypes_i\}$
                \EndFor
                \ForAll{$(a_1, \dots, a_n) \in A$} 
                    \State $\calV \gets $ \Call{Execute}{$op, (a_1, \dots, a_n)$}
                    \If{$\calV \not \in E$} \Comment{The value has not been encountered before}
                       \State $w' \gets w$
                        \color{BustleLineColor}
                        \State $s_{vo} \gets \Call{PropertySignature}{\calV, \mathcal{O}, \mathit{Props}}$ \label{alg:bustle_line_2}
                        \State $w' \gets \Call{ReweightWithModel}{M, s_{io}, s_{vo}, w}$ \label{alg:bustle_line_3}
                        \color{Black}
                        \State $E[w'] \gets E[w'] \cup \{\calV\}$
                    \EndIf
                    \If{$\calV = \mathcal{O}$}
                        \State \Return \Call{Expression}{$\calV$}
                    \EndIf
                \EndFor
            \EndFor
        \EndFor
    \end{algorithmic}
\end{algorithm}

\subsection{Benchmarks}
\label{section:benchmarks}
We evaluate \N{Bustle} on two datasets.
The first dataset is a new suite of 38 human-written benchmark tasks,
which were designed to contain a variety of tasks difficult enough to stress our system. Some tasks involve conditionals, which are not present in our other set of benchmark tasks (from SyGuS).
The search space explored by the synthesizers to solve these tasks
is quite large: on average, \N{Bustle} searches about 5 million expressions per benchmark attempt, and 1 million
expressions per successful attempt.
Most tasks have between 2 and 4 input-output pairs, though for some tasks, more than 4 pairs are needed to fully specify the semantics of 
the desired program, especially when conditionals are involved.
In each case, we gave what we felt was the number of pairs a user of 
such a system would find reasonable (though of course this is a subjective
judgment).
Three representative benchmarks are shown in Listing \ref{listing:benchmarks}.
See Appendix~\ref{app:benchmarks} for a full list.

The second dataset consists of all SyGuS programs from the
\hyperlink{https://github.com/SyGuS-Org/benchmarks/tree/master/comp/2019/PBE_SLIA_Track/euphony}{2019 PBE SLIA TRACK}
and the 
\hyperlink{https://github.com/ellisk42/ec/tree/master/PBE_Strings_Track}{2018 PBE Strings Track} whose inputs and outputs are only strings. We removed duplicate copies of problems which simply had extra examples. 
This results in 89 remaining tasks.

\begin{lstlisting}[float=*, language=Python, caption={
Three of our benchmark problems (all solved by \N{Bustle}).
}, label=listing:benchmarks]
# Compute the depth of a path, i.e., count the number of slashes
solution = "TO_TEXT(MINUS(LEN(var_0), LEN(SUBSTITUTE(var_0, \"/\", \"\"))))"
io_pairs = {"/this/is/a/path": "4", "/home": "1", "/a/b": "2"}

# Change DDMMYYYY date to MM/DD/YYYY
solution = "CONCATENATE(MID(var_0, 3, 2), \"/\", REPLACE(var_0, 3, 2, \"/\"))"
io_pairs = {"08092019": "09/08/2019", "12032020": "03/12/2020"}

# Create capitalized acronym from two words in one cell
solution =
    "UPPER(CONCATENATE(LEFT(var_0, 1), MID(var_0, ADD(FIND(\" \", var_0), 1), 1)))"
io_pairs = {"product area": "PA", "Vice president": "VP"}

\end{lstlisting}

\section{\N{Bustle}: Bottom-Up Synthesis with Learning}

\begin{figure}
\centering
\includegraphics[width=0.9\linewidth]{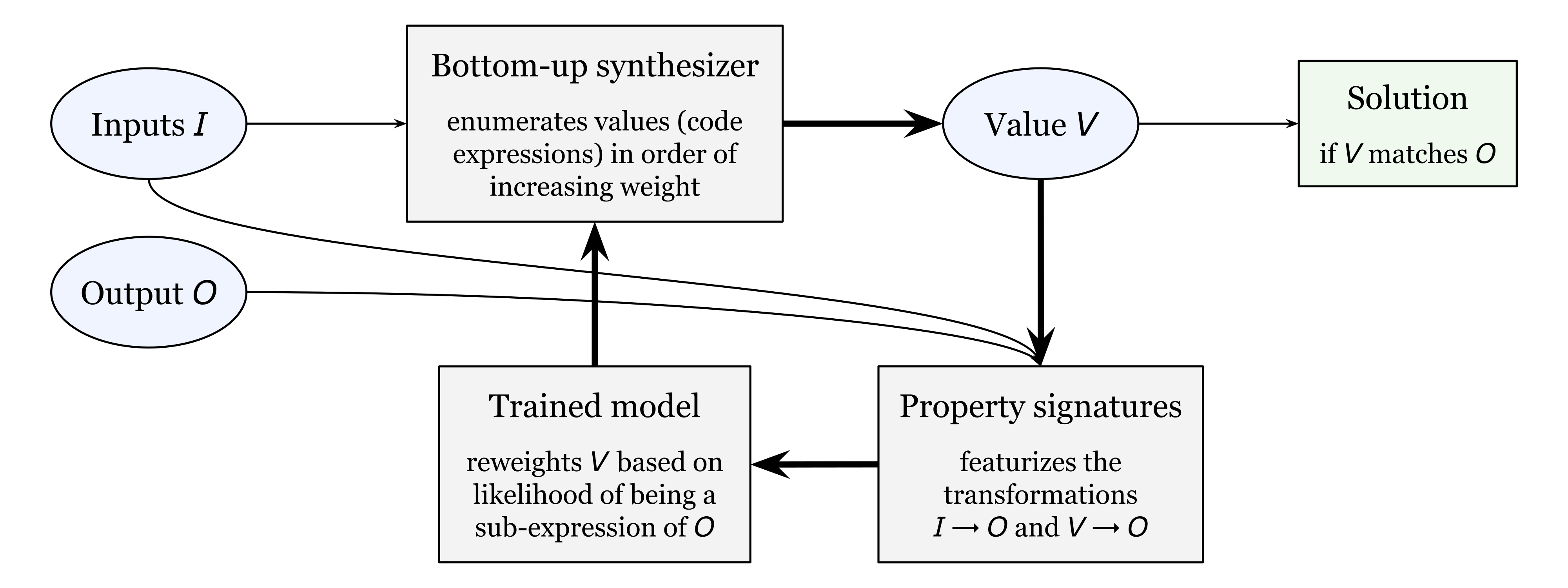}
\caption{Diagram outlining the \N{Bustle} approach. The bold arrows show the main feedback loop. The bottom-up synthesizer enumerates values (code expressions), which are featurized along with the I/O example using property signatures. The property signatures are passed to a trained model that reweights the value based on whether it appears to be a sub-expression of a solution, and the reweighted value is given back to the bottom-up synthesizer for use in enumerating larger expressions.}
\label{fig:diagram}
\end{figure}

\N{Bustle} uses a trained machine learning model to guide the bottom-up search, outlined in Figure~\ref{fig:diagram}. Suppose the synthesis task includes $n$ different input-output examples, and there are $m$ separate input variables (such that the function to synthesize has arity $m$). Recall that every value in the bottom-up search represents a code expression and also contains the $n$ results of evaluating the code expression on the $n$ examples. To guide the bottom-up search, we run a machine learning model on the results contained in intermediate values encountered during the search.

The model is a binary classifier $p(y \mid \calI, \calV, \calO)$ where $\calI$ is a set of $m$ input values (one for each input variable), $\calO$ is a single value representing the desired output, and $\calV$ is an intermediate value encountered during the bottom-up search. We want the model to predict the binary label $y=1$ if the code expression that produced $\calV$ is a sub-expression of a solution to the synthesis problem, and $y=0$
otherwise. Given these predictions, we de-prioritize sub-expressions that are unlikely to appear in the final result, which, when done correctly, dramatically speeds up the synthesizer.

\subsection{Model Architecture and Training}
\label{section:training_the_model}

Because we want the classifier to learn whether a value is intermediate
between an input and an output, the model is conditioned on \emph{two} property signatures: one from the inputs to the output,
and one from the intermediate value to the output. Recall from Section~\ref{section:property_signatures}
that a property signature is computed by applying a list of properties
to a list of input-output pairs. Thus, one signature is computed by applying all of the properties to input-output pairs, and the other is applied to intermediate value-output pairs.
A few example properties that we use include: (a) if the value $v$ and output $o$ are both strings,
is $v$ an initial substring of $o$; (b) do $v$ and $o$ have the same length; (c) does the string contain a space, and so on.
(See Appendix~\ref{app:properties} for the full list of properties.)
Then we concatenate these two vectors to obtain the model's input.
The rest of the model is straightforward.
Each element of the property signature  is either $\alltrue$, $\allfalse$, or $\mixed$.
We embed the ternary property signature
into a higher-dimensional dense vector and
then feed it into a fully connected neural network for 
binary classification. 

This model is simple, but we are only able to use such a simple model 
due to our particular design choices: our form of bottom-up search 
guarantees that all intermediate expressions can yield a value comparable
to the inputs and outputs, and property signatures can do much of the 
``representational work'' that would otherwise require a larger or more complex model.

The classifier is trained by behavior cloning on a set of training problems.
However, obtaining a training dataset is challenging.
Ideally, the training set would contain synthesis tasks that are interesting to humans, 
but such datasets can be small compared to what is needed to train deep neural networks.
Instead, we train on randomly generated synthetic data, similar to~\cite{ROBUSTFILL}. 
This choice does come with a risk of poor performance on human-written tasks due to domain mismatch
\citep{shin2019-sv},
but we show in Section~\ref{section:results} that \N{Bustle} can 
overcome this issue.

Generating the synthetic data is itself nontrivial.
Because different DSL functions have corresponding argument preconditions and invariants (e.g., several functions take integer indices which must be in range for a given string), a random sampling of DSL programs and inputs would lead to a large number of training
examples where the program cannot be applied to the sampled inputs.

Instead, we use the idea of generating data from synthesis searches, as in TF-Coder~\citep{TFCODER}.
First, we generate some random input strings to form the inputs $\calI$ (using multiple input variables and multiple examples).
From these inputs, we run bottom-up search using a dummy output, so that the search will keep generating expressions. We randomly select some generated expressions and pretend that they are outputs $\calO$ for a synthesis task. Then, for each selected expression, we randomly select one of its sub-expressions $\calV$ to form a positive training example $(\calI, \calV, \calO)$. We also randomly select another expression $\calV'$ from the search that is \emph{not} a sub-expression of $\calO$, to serve as a negative training example $(\calI, \calV', \calO)$. In our experiments, we perform 1000 searches on random inputs, select 100 values at random from each search to serve as outputs, and create one positive and one negative example for each selected value as described. Considering that a dataset of size 200,000 is not large by modern deep learning standards, training the model is a small one-time cost, completing in a few hours on CPU.

\subsection{Combining Model with Synthesis}
\label{section:synthesis_algorithm}

Incorporating the model into bottom-up synthesis is straightforward,
and can be accomplished by adding the blue lines into 
Algorithm~\ref{alg:search}. Lines~\ref{alg:bustle_line_1} 
and~\ref{alg:bustle_line_2} compute the property signatures
required for the model input, as described previously.
The main challenge is that the model produces a probability 
$p(y \mid \calI, \calV, \calO),$ but the search is organized by integer weights.
We resolve this with a simple heuristic:
at the time we generate a new value, we have the weight $w$ of the 
expression that generates it.
We discretize the model's output probability into an integer in $\delta \in \{0, \ldots, 5\}$
by binning it into six bins bounded by $[0.0, 0.1, 0.2, 0.3, 0.4, 0.6, 1.0].$
The new weight is computed from the discretized model output as $w' = w + 5-\delta.$
This function is indicated by \N{ReweightWithModel} in Algorithm~\ref{alg:search}.

A key challenge is making the model fast enough. Evaluating the model once for every 
intermediate value could cause the synthesizer to slow down so much
that the overall performance would be worse than with no model at all.
However, \N{Bustle} actually
outperforms our baselines even when measured strictly in terms of wall-clock time
(Section~\ref{section:results}).
There are several reasons for this.
First, computing property signatures for the expressions allows us to take
some of the work of representing the intermediate state out of the neural
network and to do it natively in Java (which is much faster comparatively).
Second, because a property signature is a fixed-length representation,
it can be fed into a simple feed-forward neural network, rather than requiring
a recurrent model, as would be necessary if we passed in a more complex representation such as the AST.
Third, because of this fixed-length representation, it is easy to batch
many calls to the machine learning model and process them using CPU vector instructions.
Inference calls to the machine learning model could, in principle, 
be done in parallel to the rest of the synthesizer, either on a separate CPU core
or on an accelerator, which would further improve wall-clock results, but our experiments are performed entirely on one CPU.
Due to these optimizations, computing property signatures and running the model on them accounts for only roughly 20\% of the total time spent.

\section{Experiments}
\label{section:results}

We evaluate \N{Bustle} on both datasets described in Section~\ref{section:benchmarks}.
To measure performance, we consider the number of benchmarks solved as a function of the number of candidates considered, which gives insight into how well the model can guide the search. We additionally consider benchmarks solved as a function of 
wall-clock time, which takes into account the computational cost of model inference.
 
\begin{figure}[t]
    \centering

    \begin{subfigure}[b]{\linewidth}
        \centering
        \includegraphics[width=\columnwidth]{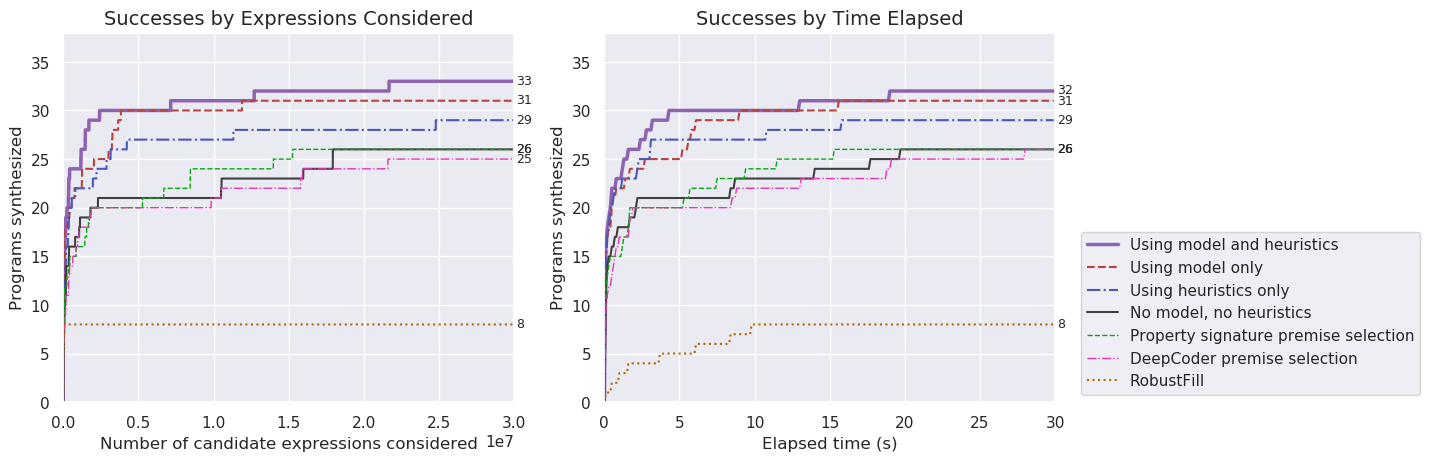}
        \caption{Results on our set of 38 new benchmarks.}
        \label{fig:progress_orig}
    \end{subfigure}

    \begin{subfigure}[b]{\linewidth}
        \centering
        \includegraphics[width=\columnwidth]{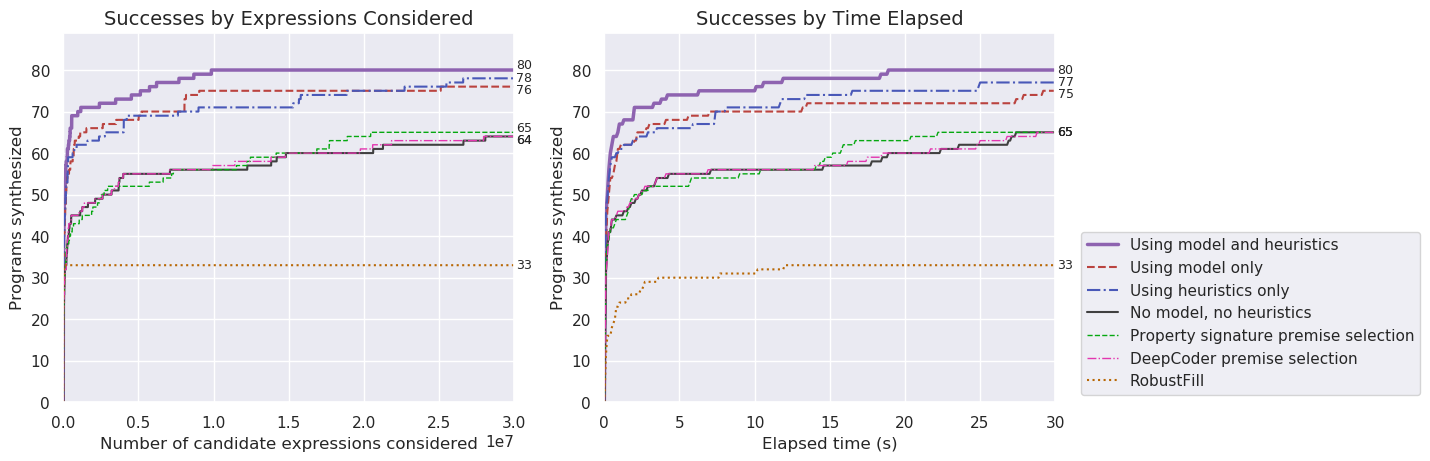}
        \caption{Results on 89 benchmarks from SyGuS.}
        \label{fig:progress_sygus}
    \end{subfigure}
    
    \caption{
    (Left) Benchmarks solved as a function of intermediate expressions considered.
    This metric makes \N{Bustle} look somewhat better than it is, 
    because it ignores slowdowns in wall-clock time, but it is still important to analyze. 
    It is invariant to engineering considerations, providing an upper bound on how well we can do in wall-clock terms through speeding up the model.
    (Right) Benchmarks solved over elapsed wall-clock time.
    \N{Bustle} still outperforms all baselines on our 38 new tasks, but not by quite as much due to time spent on model inference.}
    \label{fig:progress_plots}
\end{figure}

We compare \N{Bustle} to five other methods:
\begin{enumerate}

\item A baseline bottom-up synthesizer without machine learning, which explores expressions in order of increasing size, without any model to guide the search toward the desired output.

\item The baseline synthesizer augmented with domain-specific heuristics (substring relationships and edit distance) to reweight intermediate string values during the search.

\item An encoder-decoder model as in RobustFill~\citep{ROBUSTFILL}, which predicts a program directly from the input-output examples.
We use beam search on the decoder with a beam size of $80,000$ programs, enough to exhaust 16 GB of GPU memory. See Appendix~\ref{app:robustfill} for more details.

\item A premise selection model as in DeepCoder~\citep{DEEPCODER}, which lets us analyze whether learning within the synthesis loop is better than learning once at the beginning of the search. We train a model similar to the model trained for \N{Bustle} on the
same dataset, but instead of predicting whether an expression is a sub-expression of a solution, we
predict which operations will be used in a solution. The examples are given to the model using character-level embeddings.
Then, for each benchmark, we exclude the 2 operations
that the model predicts are the least likely.

\item A premise selection model that uses property signatures instead of character-level embeddings.

\end{enumerate}
\paragraph{Results} The results on our 38 new benchmarks are shown in Figure~\ref{fig:progress_orig}. Whether comparing by the number of expressions (left) or the wall-clock time (right), \N{Bustle} (red-dash) performs quite well, solving 31 tasks within 30 million candidate expressions or 30 seconds.
It outperforms all other methods besides one that uses \textit{both} the model and the heuristics (purple). In particular, \N{Bustle} outperforms the domain-specific heuristics, even though the heuristics are much faster to execute compared to running model inference.
Furthermore, when using both the model and heuristics (purple), the synthesizer performs the best overall.
This indicates that although learning in the loop outperforms domain-specific heuristics, the two approaches can be combined to achieve better performance than either alone.

Results on SyGuS benchmarks in Figure~\ref{fig:progress_sygus} were broadly similar, with one exception: 
the handwritten heuristics perform slightly better than (heuristic-free)
\N{Bustle}.
There are a few important caveats to this, however.
First, the SyGuS problems are slightly different from the kinds of problems in our 38 new benchmark tasks and the training data, e.g., some SyGuS problems have an incredibly large number of examples or use string-manipulation functionality outside our DSL.
Second, our heuristics substantially outperform all other baselines, 
which suggests that they are strong heuristics to begin with.
Third, we still see a substantial improvement by \textit{combining} the \N{Bustle}
model with heuristics, so the best performing algorithm does indeed use \N{Bustle}.

\N{Bustle} outperforms both DeepCoder-style premise selection methods (green and pink).
Premise selection allows some tasks to be solved faster, but it does not lead to more tasks being solved overall compared to the no-model baseline (black). This is evidence that learning in the loop is important to guide the search as it happens, and one step of learning at the beginning of search is not as effective.
We furthermore observe that using property signatures (green) leads to better performance than not (pink), since they can help the models be more robust to train-test distribution shift.

\N{Bustle} also outperforms RobustFill (orange).
Relatively speaking, RobustFill performs better on the SyGuS tasks than our new tasks, which may indicate that some of our new tasks are more difficult due to the use of conditionals.
Overall, RobustFill does not perform well, possibly because the end-to-end neural approach is less robust to the train-test distribution shift, and because its complex model cannot predict programs as quickly as \N{Bustle}'s fast combinatorial search.

We conduct two additional analyses to better understand the performance of \N{Bustle}.
First, we investigate the predictions of the model when it is run 
on the intermediate values encountered during synthesis of the 
human-written benchmarks.
We generate histograms for the model's predictions on expressions
that do appear in the solution and expressions that do not appear in 
the solution, for all benchmarks that were solved.
Predictions for true sub-expressions skew positive and predictions for negative sub-expressions skew negative. This provides further evidence that our model generalizes well to human benchmarks,
despite the domain mismatch to the synthetic data used in training.
The full results are in Appendix~\ref{app:histogram}.
Finally, we determined that all but one of the benchmarks solved by the baseline (no model, no heuristics) were also solved by \N{Bustle}, across both sets of benchmarks.

\section{Related Work}

For surveys on program synthesis
and machine learning for software engineering, see
\citet{THREEPILLARS,ARMANDOCOURSE,RISHABHSURVEY,BIGCODE}.
A well-known synthesizer for spreadsheet programs is FlashFill \citep{FLASHFILL}, based on
Version Space Algebras (VSAs)
which are powerful and fast. However, VSAs only apply to restricted DSLs,
e.g., the top-most function 
in the program must be \T{Concat}, which allows it to perform efficient divide and conquer style search. 
Our technique has no such restrictions.

Early work on machine learning for program synthesis includes 
DeepCoder \citep{DEEPCODER}, which uses a learned model to select 
once at the beginning of search. Although this idea
is pragmatic, the disadvantage is that
once the search has started, the model can give no further feedback.
\cite{SIGNATURES} use property signatures within a DeepCoder-style model for premise selection.

One can also train a machine learning model to emit whole 
programs token-by-token using an encoder-decoder neural
architecture \citep{bunel2018leveraging,ROBUSTFILL,parisotto2017neuro}, but this approach does not have the ability to inspect outputs of intermediate programs.
Previous work has also considered using learning within syntax guided search over programs
\citep{Yin2017-my,EUPHONY}, but because these methods are top-down, it is much more
difficult to guide them by execution information, since partial programs will have holes.
Finally, \citet{INFERSKETCHES} learns to emit a partial program and fill
in the holes with a symbolic search.

The most closely related work to ours is \emph{neural execution-guided
synthesis}, which like \N{Bustle} uses values produced by intermediate programs within
a neural network.
\citet{GARBAGECOLLECTOR} process intermediate values of a program using a 
neural network for a small, straight-line DSL, but they do not use the model
to evaluate intermediate programs.
Another approach is to rewrite a programming language so that it can
be evaluated ``left-to-right'', allowing values to be used to prioritize the search in
an actor-critic framework \citep{REPL}.
Similarly, \citet{ChenLS19} use intermediate values while synthesizing a 
program using a neural encoder-decoder model, but again this work proceeds in a variant
of left-to-right search that is modified to handle conditionals and loops.
None of these approaches exploit our main insight, which is that bottom-up search
allows the model to prioritize and combine small programs that solve
different subtasks.

\emph{Learning to search} has been an active area in machine learning,
especially in imitation learning \citep{Daume2009-lm,Ross2011-li,Chang2015-od}. 
Combining more sophisticated imitation learning strategies into \N{Bustle}
is an interesting direction for future work.

\section{Conclusion}

We introduced \N{Bustle}, a technique for using machine learning to guide bottom-up
search for program synthesis.
\N{Bustle} exploits the fact that bottom-up search makes it easy to evaluate
partial programs, and it uses machine learning to predict the likelihood that a given
intermediate value is a sub-expression of the desired solution.
We have shown that \N{Bustle} improves over various baselines, including recent deep-learning-based program synthesis approaches (DeepCoder and RobustFill), on two challenging benchmark suites of string-manipulation tasks, in terms of candidate programs considered as well as wall-clock time.
In fact, showing that learning-in-the-loop can be made fast enough for
program synthesis is perhaps the major contribution of this work. The idea
of learning-in-the-loop, though novel as far as we are aware, is relatively 
obvious, but through this work we learned that it can be efficient enough to provide time speedups overall.

\subsubsection*{Acknowledgments}
The authors thank Sam Schoenholz and the anonymous conference reviewers for their helpful reviews and comments on our paper.

\bibliography{iclr2021_conference}
\bibliographystyle{iclr2021_conference}

\newpage
\appendix

\section{Expanded Description of DSL}
\label{app:dsl}
Our DSL allows for nesting and compositions of common string transformation functions. These functions include string concatenation (\T{Concat});
returning a substring at the beginning (\T{Left}), middle (\T{Substr}),
or right (\T{Right}) of a string;
replacing a substring of one string, indicated by start and end position,
with another string (\T{Replace}); 
removing white space from the beginning and ending of a string (\T{Trim});
concatenating a string with itself a specified number of times (\T{Repeat});
substituting the first $k$ occurrences of a substring with another (\T{Substitute});
converting an integer to a string (\T{ToText});
and converting a string to \T{LowerCase}, \T{UpperCase}, or
every word capitalized (\T{ProperCase}).
Integer functions include arithmetic, returning the index of the 
first occurrence
of a substring within a string (\T{Find}), and string length (\T{Len}).
We also have some functions either consuming or producing booleans
(\T{If}, \T{Equals}, \T{GreaterThan}, \T{GreaterThanOrEqualTo}).
Finally, a few commonly useful string and integer constants
are included.

\section{List of Benchmark Programs}
\label{app:benchmarks}
Here we show each of our 38 human-written benchmark problems, and a possible solution
written in a DSL that is a superset of the DSL used by our synthesizer.
We have separated them into 
Listing \ref{listing:benchmarks1}
Listing \ref{listing:benchmarks2} for space reasons.
Note that the synthesizer can and does solve problems with programs different than 
the programs given here, and that it does not solve all of the problems.

\section{Listing of Properties Used}
\label{app:properties}
\N{Bustle} computes two types of property signatures:
the signature involving the inputs and the outputs,
and the signature involving the intermediate state and the outputs.
In this paper, the inputs and outputs are always strings, but the 
intermediate state may be an integer, a string, or a boolean.
In an abstract sense, a property acts on an input and an output,
but some properties will simply ignore the input, 
and so we implement those as functions with only one argument.
Thus, we have six types of properties in principle:

\begin{itemize}
\item properties acting on a single string (Listing \ref{listing:singlestringproperties}).
\item properties acting on a single integer (Listing \ref{listing:singleintegerproperties}).
\item properties acting on a single boolean (there is only one of these).
\item properties acting on a string and the output string
(Listing \ref{listing:stringstringproperties}).
\item properties acting on an integer and the output string
(Listing \ref{listing:integerstringproperties}).
\item properties acting on a boolean and the output string (we don't actually
use any of these presently).
\end{itemize}

For a program with multiple inputs, we simply loop over all the inputs and the output,
computing all of the relevant types of properties for each.
For example, a program taking two string inputs and yielding one string
output will have single-argument string properties for the output and 
two sets of double-argument string properties, one for each input.
We fix a maximum number of inputs and pad the signatures so that they are all the same size.

\section{More implementation details about RobustFill baseline}
\label{app:robustfill}

To make the vanilla model proposed in~\citet{ROBUSTFILL} work on our benchmarks, we have made the following necessary changes: 
\begin{itemize}
    \item As the vanilla model only allows single input per each input-output pair, we here concatenate the variable number of inputs with a special separator token.
    \item Following \citet{ROBUSTFILL}, the vocabulary of input-output examples are constructed at the character level, as there are numerous integer and string literals in our benchmark and training tasks.
    \item The desired programs may use string constants which may depend on the particular input-output examples. In addition to predicting program tokens, RobustFill may also need to predict the string constants character-by-character when necessary.
    \item There could be out-of-vocabulary characters in the input-output examples, as the test benchmarks have a different data distribution than synthetic training examples. In this case, we replace these characters with spaces.
\end{itemize}

We use the same training data to train the RobustFill model. We first retain 10\% of the training samples for validation, and identify the best number of training epochs. Then we retrain the model using full data for that many epochs. The batch size we use is 1024, with fixed learning rate 1e-3. When decoding the program, we use a 3-layer LSTM with embedding size of 512.

During inference, we use beam search to obtain the most likely $M$ programs, following~\citet{ROBUSTFILL}. On a single GPU with 16 GB of memory, the maximum beam size we can use is $M = 80,000$, which is already several magnitudes larger than $1000$ (the largest beam size used in original paper). It takes roughly 25 seconds to perform this beam search, which is conveniently close to the 30 second time limit in our experiments.

\section{Analysis of Model Predictions}
\label{app:histogram}
\begin{figure}[t]
\begin{center}
\includegraphics[width=0.7\columnwidth]{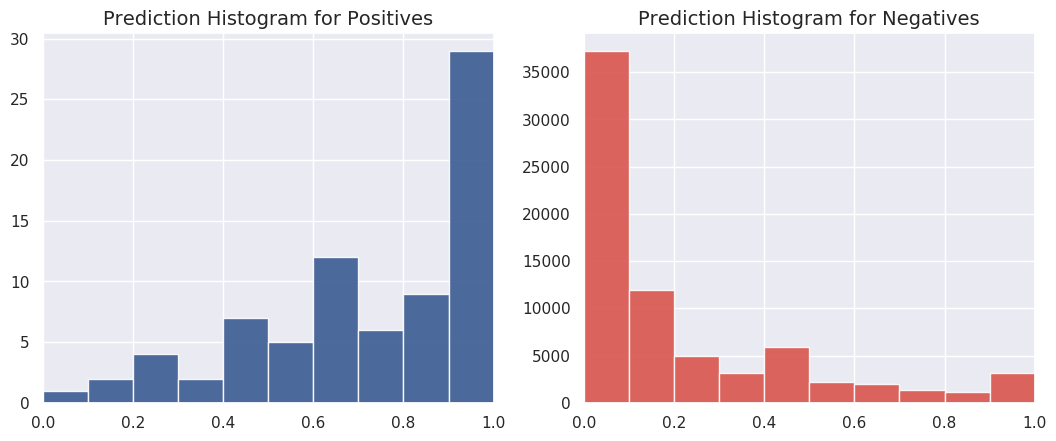}
\caption{
Histograms of model predictions for expressions seen while solving benchmarks.
(Left) for expressions that were sub-expressions of a solution, the majority received predictions close to 1, showing that the model can identify the correct expressions to prioritize during search.
(Right) for expressions that were not sub-expressions of a solution, predictions skewed close to 0.}
\label{fig:prediction_histograms}
\end{center}
\end{figure}

We investigate the predictions of the model when it is run 
on the intermediate values actually encountered during synthesis of the 
human-written benchmarks.
We compute separate histograms for the model's predictions on expressions
that do appear in the solution and expressions that do not appear in 
the solution, for all benchmarks that were solved.
Predictions for true sub-expressions skew positive and predictions for negative sub-expressions skew negative. This provides further evidence that our model generalizes well to human benchmarks, despite the domain mismatch to the synthetic data used in training. See Figure~\ref{fig:prediction_histograms}.

\begin{lstlisting}[float=*, language=Java, caption={
Potential solutions for our benchmarks, along with comments describing the 
semantics of the solution.
}, label=listing:benchmarks1]
// add decimal point if not present
IF(ISERROR(FIND(".", var_0)), CONCATENATE(var_0, ".0"), var_0)

// add plus sign to positive integers
IF(EXACT(LEFT(var_0, 1), "-"), var_0, CONCATENATE("+", var_0))

// append AM or PM to the hour depending on if it's morning
CONCATENATE(LEFT(var_0, MINUS(FIND(":",var_0), 1)), IF(EXACT(var_1, "morning"), " AM", " PM"))
                                               
// fix capitalization of city and state
CONCATENATE(LEFT(PROPER(var_0), MINUS(LEN(var_0), 1)), UPPER(RIGHT(var_0, 1)))

// capitalize the first word and lowercase the rest
REPLACE(LOWER(var_0), 1, 1, UPPER(LEFT(var_0, 1)))

// whether the first string contains the second
TO_TEXT(ISNUMBER(FIND(var_1, var_0))) 

// whether the first string contains the second, ignoring case 
TO_TEXT(ISNUMBER(FIND(LOWER(var_1), LOWER(var_0))))
                                           
// count the number of times the second string appears in the first
TO_TEXT(DIVIDE(MINUS(LEN(var_0), LEN(SUBSTITUTE(var_0, var_1, ""))), LEN(var_1)))

// create email address from name and company  
LOWER(CONCATENATE(LEFT(var_0, 1), var_1, "@", var_2, ".com"))

// change DDMMYYYY date to MM/DD/YYYY
CONCATENATE(MID(var_0, 3, 2), "/", REPLACE(var_0, 3, 2, "/"))         

// change YYYY-MM-DD date to YYYY/MM/DD
SUBSTITUTE(var_0, "-", "/")

// change YYYY-MM-DD date to MM/DD
SUBSTITUTE(RIGHT(var_0, 5), "-", "/")

// extract the part of a URL between the 2nd and 3rd slash
MID(var_0, ADD(FIND("//", var_0), 2), MINUS(MINUS(FIND("/", var_0, 9), FIND("/", var_0)), 2))

// extract the part of a URL starting from the 3rd slash
RIGHT(var_0, ADD(1, MINUS(LEN(var_0), FIND("/", var_0, ADD(FIND("//", var_0), 2)))))

// get first name from second column
LEFT(var_1, MINUS(FIND(" ", var_1), 1))

// whether the string is lowercase
IF(EXACT(var_0, LOWER(var_0)), "true", "false")

// get last name from first column
RIGHT(var_0, MINUS(LEN(var_0), FIND(" ", var_0)))

// output "Completed" if 100%, "Not Yet Started" if 0%, "In Progress" otherwise
IF(var_0="100%", "Completed", IF(var_0="0%", "Not Yet Started", "In Progress"))

// enclose negative numbers in parentheses
IF(EXACT(LEFT(var_0, 1), "-"), CONCATENATE(SUBSTITUTE(var_0, "-", "("), ")"), var_0)

\end{lstlisting}

\begin{lstlisting}[float=*, language=Java, caption={
Potential solutions for our benchmarks, along with comments describing the 
semantics of the solution.
}, label=listing:benchmarks2]
// pad text with spaces to a given width
CONCATENATE(REPT(" ", MINUS(VALUE(var_1), LEN(var_0))), var_0)

// pad number with 0 to width 5
CONCATENATE(REPT("0", MINUS(5, LEN(var_0))), var_0)

// the depth of a path, i.e., count the number of /
TO_TEXT(MINUS(LEN(var_0), LEN(SUBSTITUTE(var_0, "/", ""))))

// extract the rest of a word given a prefix
RIGHT(var_0, MINUS(LEN(var_0), LEN(var_1)))

// prepend Mr. to last name
CONCATENATE("Mr. ", RIGHT(var_0, MINUS(LEN(var_0), FIND(" ", var_0))))

// prepend Mr. or Ms. to last name depending on gender
CONCATENATE(IF(EXACT(var_1, "male"), "Mr. ", "Ms. "), 
            RIGHT(var_0, MINUS(LEN(var_0), FIND(" ", var_0))))

// remove leading and trailing spaces and tabs, and lowercase
TRIM(LOWER(var_0))

// replace <COMPANY> in a string with a given company name
SUBSTITUTE(var_0, "<COMPANY>", var_1)

// replace com with org
SUBSTITUTE(var_0, "com", "org", 1)

// select the first string, or the second if the first is NONE
IF(EXACT(var_0, "NONE"), var_1, var_0)

// select the longer of 2 strings, defaulting to the first if equal length
IF(GT(LEN(var_1), LEN(var_0)), var_1, var_0)

// whether the two strings are exactly equal, yes or no
IF(EXACT(var_0, var_1), "yes", "no")

// whether the two strings are exactly equal ignoring case, yes or no
IF(EXACT(LOWER(var_0), LOWER(var_1)), "yes", "no")

// length of string
TO_TEXT(LEN(var_0))

// extract the rest of a word given a suffix
LEFT(var_0, MINUS(LEN(var_0), LEN(var_1)))

// swap the case of a string that is entirely uppercase or lowercase
IF(EXACT(var_0, LOWER(var_0)), UPPER(var_0), LOWER(var_0))

// truncate and add ... if longer than 15 characters 
IF(GT(LEN(var_0), 15), CONCATENATE(LEFT(var_0, 15), "..."), var_0)

// create acronym from two words in one cell
CONCATENATE(LEFT(var_0, 1), MID(var_0, ADD(FIND(" ", var_0), 1), 1))

// create capitalized acronym from two words in one cell
UPPER(CONCATENATE(LEFT(var_0, 1), MID(var_0, ADD(FIND(" ", var_0), 1), 1)))
\end{lstlisting}

\begin{lstlisting}[float=t, language=Java, caption={
Java code for all Properties acting on single Strings.
}, label=listing:singlestringproperties]
str.isEmpty()               // is empty?
str.length() == 1           // is single char?
str.length() <= 5           // is short string?
str.equals(lower)           // is lowercase?
str.equals(upper)           // is uppercase?
str.contains(" ")           // contains space?
str.contains(",")           // contains comma?
str.contains(".")           // contains period?
str.contains("-")           // contains dash?
str.contains("/")           // contains slash?
str.matches(".*\\d.*")      // contains digits?
str.matches("\\d+")         // only digits?
str.matches(".*[a-zA-Z].*") // contains letters?
str.matches("[a-zA-Z]+")    // only letters?
\end{lstlisting}

\begin{lstlisting}[float=t, language=Java, caption={
Java code for all Properties acting on single Integers.
}, label=listing:singleintegerproperties]
integer == 0                // is zero?
integer == 1                // is one?
integer == 2                // is two?
integer < 0                 // is negative?
0 < integer && integer <= 3 // is small integer?
3 < integer && integer <= 9 // is medium integer?
9 < integer                 // is large integer?
\end{lstlisting}

\begin{lstlisting}[float=t, language=Java, caption={
Java code for all Properties acting on a String and the output String.
}, label=listing:stringstringproperties]
outputStr.contains(str)            // output contains input?
outputStr.startsWith(str)          // output starts with input?
outputStr.endsWith(str)            // output ends with input?
str.contains(outputStr)            // input contains output?
str.startsWith(outputStr)          // input starts with output?
str.endsWith(outputStr)            // input ends with output?
outputStrLower.contains(lower)     // output contains input ignoring case?
outputStrLower.startsWith(lower)   // output starts with input ignoring case?
outputStrLower.endsWith(lower)     // output ends with input ignoring case?
lower.contains(outputStrLower)     // input contains output ignoring case?
lower.startsWith(outputStrLower)   // input starts with output ignoring case?
lower.endsWith(outputStrLower)     // input ends with output ignoring case?
str.equals(outputStr)              // input equals output?
lower.equals(outputStrLower)       // input equals output ignoring case?
str.length() == outputStr.length() // input same length as output?
str.length() < outputStr.length()  // input shorter than output?
str.length() > outputStr.length()  // input longer than output?
\end{lstlisting}

\begin{lstlisting}[float=t, language=Java, caption={
Java code for all Properties acting on an Integer and the output String.
}, label=listing:integerstringproperties]
integer < outputStr.length()                // is less than output length?
integer <= outputStr.length()               // is less or equal to output length?
integer == outputStr.length()               // is equal to output length?
integer >= outputStr.length()               // is greater or equal to output length?
integer > outputStr.length()                // is greater than output length?
Math.abs(integer - outputStr.length()) <= 1 // is very close to output length?
Math.abs(integer - outputStr.length()) <= 3 // is close to output length?
\end{lstlisting}

\end{document}